\documentclass[aps,pra,twocolumn,showpacs,amsmath]{revtex4-1}
\usepackage{amssymb,amsmath}
\usepackage{mathrsfs}
\usepackage{graphicx}
\usepackage{float}
\usepackage{txfonts}
\usepackage[position=t,singlelinecheck=off,
caption=false]{subfig}
\usepackage[normalem]{ulem}
\usepackage{color}
\usepackage{bm}
\usepackage{xcolor}
\usepackage{pdfcomment}
\usepackage{graphicx}
\usepackage{dcolumn}
\usepackage{bm}
\usepackage{bbm}
\usepackage{soul}
\setstcolor{red}

\usepackage[T1]{fontenc}
\usepackage[latin9]{inputenc}
\usepackage{setspace}
\usepackage{esint}

\begin{document}

\preprint{APS/123-QED}

\title{Recipe for creating an arbitrary number of Floquet chiral edge states}

\author{Longwen \surname{Zhou}}\email{zhoulw13@u.nus.edu}
\affiliation{Department of Physics, National University of Singapore, Singapore 117551, Republic of Singapore}

\author{Jiangbin \surname{Gong}}\email{phygj@nus.edu.sg}
\affiliation{Department of Physics, National University of Singapore, Singapore 117551, Republic of Singapore}

\date{\today}

\begin{abstract}
Floquet states of periodically driven systems could exhibit rich topological
properties. Many of them are absent in their static counterparts.
One such example is the chiral edge states in anomalous Floquet topological
insulators, whose description requires a new topological invariant
and a novel type of bulk-edge correspondence. In this work, we propose
a prototypical quenched lattice model, whose two Floquet bands could
exchange their Chern numbers periodically and alternatively via touching at quasienergies
$0$ and $\pi$ under the change of a single system parameter. This
process in principle allows the generation of as many Floquet
chiral edge states as possible in a highly controllable manner. The
quantized transmission of these edge states are extracted from the
Floquet scattering matrix of the system. The flexibility in controlling
the number of topological edge channels provided by our scheme
could serve as a starting point for the engineering of robust Floquet
transport devices.
\end{abstract}

\maketitle

\section{Introduction}\label{sec:Int}
Floquet states of matter emerge from systems that are modulated periodically in time~\cite{ColdAtomFTP,ColdAtomFTP2,PhotonFTP0,PhotonFTP1,PhotonFTP2,PhononFTP,PhononFTP2}.
They possess intriguing transport and topological properties~\cite{DKRGong,OkaPRB2009,LindnerNP2011,DahlhausPRB2011,KitagawaPRB2011,DerekPRL2012,HailongPRE2013,TongPRB2013,LeonPRL2013,CayssolRRL2013,ThakurathiRRB2013,DerekPRB2014,GrushinPRL2014,Wang2014,ZhouEPJB2014,TitumPRL2015,XiongPRB2016,DLossPRLPRB,KR4}, many of which are characterized by new
types of topological invariants, classification schemes and bulk-edge
correspondence that goes beyond any time-independent descriptions~\cite{KitagawaPRB2010,JiangPRL2011,KunduPRL2013,Bomantara2016,ZhaoErH2014,ReichlPRA2014,LeonPRB2014,AnomalousESPRX,FulgaPRB2016,YapPRB2017,YapMajorana2017,ZhouPRB2016,AsbothSTF,NathanNJP2015,ClassificationFTP1,ClassificationFTP2}.

One example is the anomalous chiral edge states in Floquet topological
insulators~\cite{AnomalousESPRX}. These states could traverse the Floquet gap at $\pi$-quasienergy,
connecting the top of the highest and the bottom of the lowest bulk bands in the Floquet quasienergy Brillouin zone. They are characterized by a topological winding number defined at
a given quasienergy within the gap, which is obtained by integrating both
quasimomenta over the system's Brillouin zone and time over a driving
period. With these anomalous edge states, the difference of winding
numbers in the gaps above and below a Floquet band gives its Chern
number, but the summation of Chern numbers below a Floquet gap cannot
tell the number of chiral edge states traversing it from bottom to top. This leads to
the identification of a new bulk-edge relation unique to Floquet systems~\cite{AnomalousESPRX}.
The anomalous chiral edge states have also been used to achieve quantized
non-adiabatic pumping in both clean and disordered samples~\cite{AnomalousESPRX,ZhouPRB2016}.

To date, anomalous Floquet chiral edge states have been
observed in photonic~\cite{PhotonFTP2} and acoustic~\cite{PhononFTP2}
systems. However, the experimentally realized models
support only a single pair of chiral edge states in each gap, limiting
its potential in the study of possible Floquet phases with many chiral
edge states and large winding numbers. Floquet topological phases with many chiral edge states could not only admit rich topological structures~\cite{DerekPRL2012,ZhouEPJB2014,DerekPRB2014,XiongPRB2016}, but also be useful in realizing
Floquet transport devices with a large number of topologically protected
channels along the edge~\cite{YapPRB2017,YapMajorana2017}. In this work, we propose a simple scheme
to generate any given number of chiral edge states
in a Floquet system, and demonstrate our scheme in a prototypical
quenched lattice model. The two-terminal transport of the model is
also studied using the Floquet scattering matrix approach.

\section{Recipe for creating many chiral edge states}\label{sec:BBC}
\begin{figure}
	\centering
	\includegraphics[width=.48\textwidth]{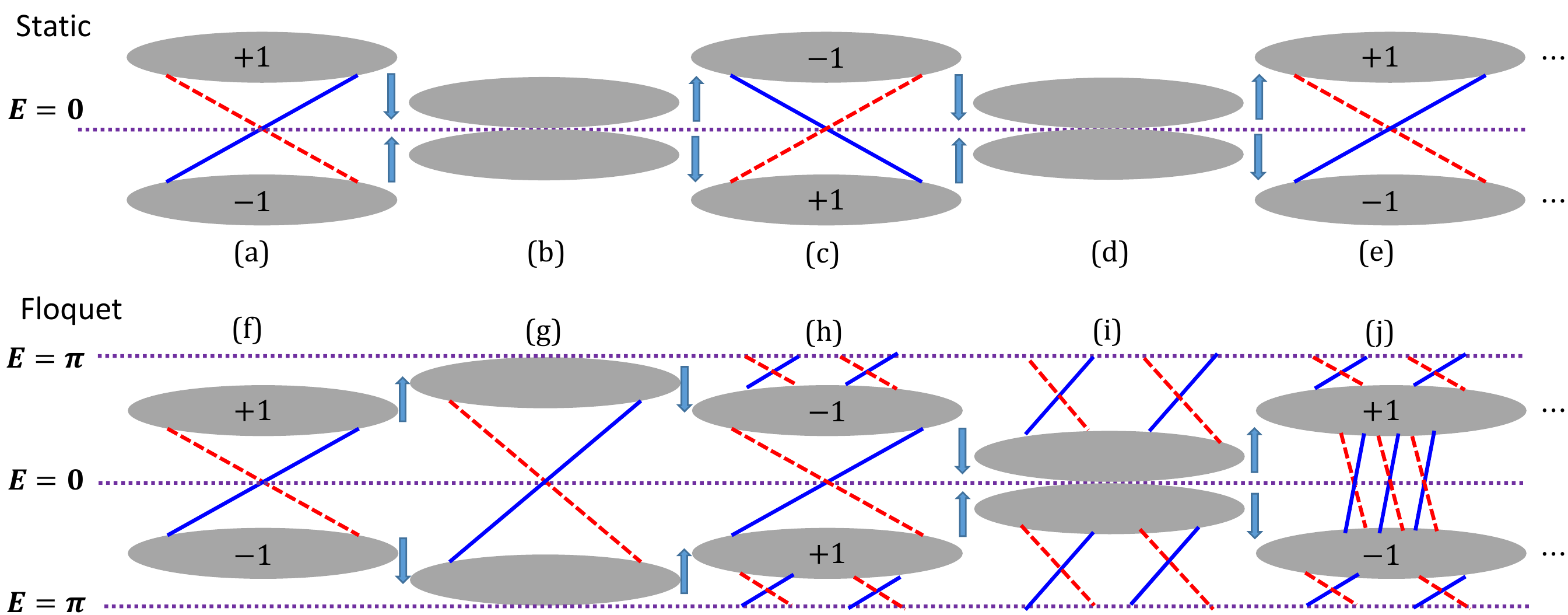}
	\caption{(color online) The scheme of generating many chiral
		edge states by periodically exchanging Floquet band Chern numbers,
		alternatively via band touchings at $0$ and $\pi$ quasienergies under the change of
		a system parameter (from left to right in each case). For demonstration
		purpose, a two-band Floquet system is presented, whereas the scheme
		is also applicable to multiple-band systems. In both cases, shaded
		regions represent bulk bands, and blue solid (red dashed) lines denote
		states localized around the left (right) edge of the lattice. Panels
		(a-e) represent the process in a static system: after two permutations
		of Chern numbers via band touchings, the system goes back to its initial topological
		phase and no new chiral edge states appear. Panels (f-j) illustrate
		the process in a Floquet system: after two permutations of band Chern
		numbers, alternatively at $\pi$ and $0$ quasienergies via band touchings, the system
		reaches a new topological phase with the same Chern number as before
		but possessing more chiral edge states traversing both Floquet gaps
		around $0$ and $\pi$ quasienergies.}
	\label{fig:Scheme}
\end{figure}

In this section, we introduce our Floquet engineering scheme, which
in principle allows the generation of arbitrarily many
chiral edge states in a well-controlled manner. An illustration of
the process is shown in Fig.~\ref{fig:Scheme}. For simplicity, we consider a two-band
insulator with band Chern numbers $\pm1$, and assume that under the
increase of a system parameter, the two bands exchange their Chern
numbers every time when they touch with each other and re-separate. 

It is instructive to compare the situations in a static and a Floquet
system. In a static system, the two bands (shaded areas in Fig.~\ref{fig:Scheme})
can only touch by closing the gap around energy $E=0$, as shown in 
Fig.~\ref{fig:Scheme}(b). Due to the bulk-edge correspondence, the two chiral edge
bands denoted by blue solid and red dashed lines in Fig.~\ref{fig:Scheme}(a) will
exchange their chiralities after the gap reopens around $E=0$ as shown in Fig.~\ref{fig:Scheme}(c), whereas the net number of chiral edge states in the gap does not change during
the transition. After a second topological phase transition, in which
the two bands exchange their Chern numbers again~[Fig.~\ref{fig:Scheme}(d)], the system will
go back to its initial topological phase with the same number of chiral
edge states~[Fig.~\ref{fig:Scheme}(e)], and the story ends here.

The situation in a Floquet system, however, can be much richer. Due
to the periodicity of Floquet quasienergy $E$, a two-band Floquet
insulator has two gaps at $0$- and $\pi$-quasienergies. Therefore,
the two Floquet bands can exchange their Chern numbers by touching
at either quasienergy $0$ or $\pi$. Now if the increase of a system
parameter could result in the closure of Floquet gaps at quasienergies
$0$ and $\pi$ \emph{periodically} and \emph{alternatively}, more
and more chiral edges should emerge in both gaps in order to compensate
for the exchange of Floquet band Chern numbers during each topological
phase transitions. 

One example of such a process is sketched in Fig.~\ref{fig:Scheme}(f-j) (with the increase
of a system parameter from left to right panels). At the starting
point~{[}Fig.~\ref{fig:Scheme}(f){]}, we have a Floquet insulator
with bulk Chern numbers $\pm1$ and two chiral edge bands crossing
the Floquet gap at quasienergy $0$. With the increase of a system
parameter, the two bulk bands gradually shift upward and downward,
respectively, until exchanging their Chern numbers upon touching at quasienergy
$\pi$~[Fig.~\ref{fig:Scheme}(g)]. But since the number of chiral edge states at quasienergy
$0$ cannot change during this process, two extra pairs of anomalous
chiral edge bands crossing the $\pi$-quasienergy
gap must appear after the transition. The resulting band structure is shown in
Fig.~\ref{fig:Scheme}(h), where we have one pair (two pairs)
of normal (anomalous) chiral edge bands in the $0$-($\pi$-) quasienergy
gap. With further increasing of the system parameter, the two bands
``kiss'' again and exchange their Chern numbers at quasienergy $0$~[Fig.~\ref{fig:Scheme}(i)].
This time, the number of anomalous chiral edge states at quasienergy
$\pi$ cannot change, and therefore the number of chiral edge bands
crossing quasienergy $0$ must increase by $2$ after the transition.
The resulting Floquet band structure is shown in
Fig.~\ref{fig:Scheme}(j). Though sharing the same Chern numbers with the initial
topological phase~[Fig.~\ref{fig:Scheme}(f)], the final system possesses two more pairs of chiral
edge bands in both $0$- and $\pi$-quasienergy gaps, and therefore
should be characterized by larger topological winding numbers~\cite{AnomalousESPRX}. It
is not hard to image that if this process could continue periodically
with the increase of the system parameter, we would in principle reach
a topological phase with arbitrarily large winding numbers, and therefore
obtaining as many chiral edge states as possible in both Floquet gaps. 

The question is how complicated a system should be to realize such
an intriguing process. In the following section, we will introduce
a periodically quenched two-dimensional ($2$d) lattice model with
only nearest neighbor hoppings. It will be shown that this
simple model realizes exactly the sequence of topological phase transitions
described in this section, which is accompanied by a monotonic increasing
of the number of Floquet chiral edge states under the increase
of just a single hopping parameter of the lattice.

\section{Prototypical model: a periodically quenched lattice}\label{sec:Model}
\begin{figure}
	\centering
	\includegraphics[width=.45\textwidth]{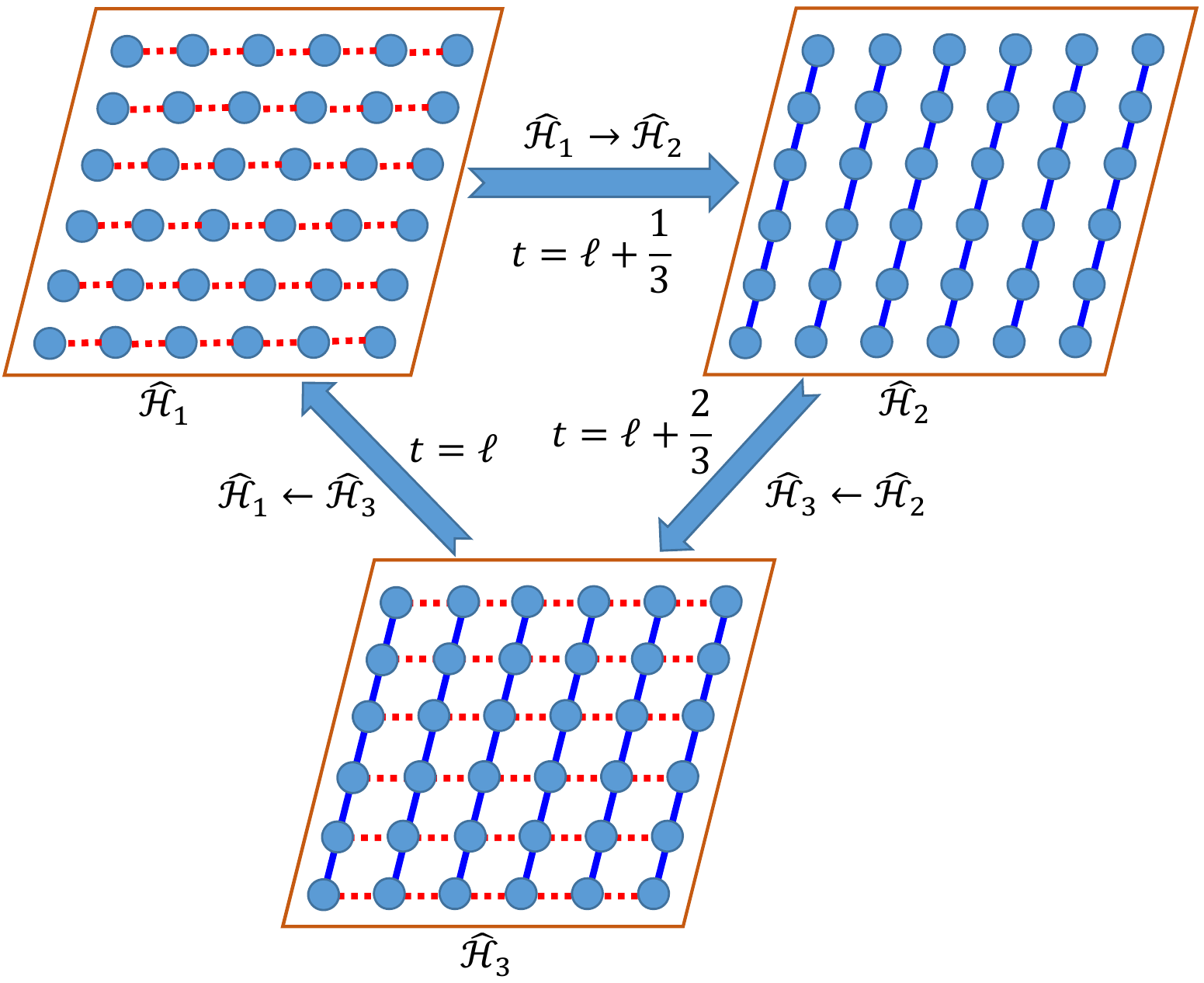}
	\caption{(color online) A sketch of the quenched lattice model
		in the $\ell$'s driving period. The same sequence of quenches is
		applied to the lattice for every driving period $\ell\in\mathbb{Z}$.
		Each shaded dot represents a unit cell with two internal degrees of
		freedom (sublattice or spin). At $t=\ell$, the system Hamiltonian
		is quenched to $\hat{{\cal H}}_{1}$, for which there are only NN
		hoppings (red dashed lines) along $x$ (horizontal) direction of the
		lattice. At $t=\ell+\frac{1}{3}$, the system Hamiltonian is switched
		to $\hat{{\cal H}}_{2}$, for which there are only NN hoppings (blue
		solid lines) along $y$ (vertical) direction of the lattice. At $t=\ell+\frac{2}{3}$,
		the system's Hamiltonian is quenched to $\hat{{\cal H}}_{3}$, for
		which NN hoppings along both $x$ and $y$ directions of the lattice
		are switched on, and an energy bias between two internal degrees of
		freedom is applied within each unit cell.}
	\label{fig:Geometry}
\end{figure}

Our model contains noninteracting particles in a $2$d square lattice,
with $2$ degrees of freedom (sublattice or spin) in each unit cell.
The nearest neighbor (NN) hopping amplitude and onsite potential of
the lattice are periodically modulated in time. In each driving period,
the system is subjected to a sequence of three quenches, as sketched
in Fig.~\ref{fig:Geometry}. The dynamics of the system following each quench is described
by the Schrodinger equation $i\frac{d}{dt}|\psi(t)\rangle=\hat{{\cal H}}|\psi(t)\rangle$,
with the Hamiltonian
\begin{equation}
\hat{{\cal H}}=\hat{{\cal H}}_{1}=\frac{3J_{1}}{2i}\sum_{n_{x},n_{y}}(|n_{x},n_{y}\rangle\langle n_{x}+1,n_{y}|-{\rm h.c.})\otimes\sigma_{x}
\end{equation}
for $t\in\left[\ell,\ell+\frac{1}{3}\right)$,
\begin{equation}
\hat{{\cal H}}=\hat{{\cal H}}_{2}=\frac{3J_{2}}{2i}\sum_{n_{x},n_{y}}(|n_{x},n_{y}\rangle\langle n_{x},n_{y}+1|-{\rm h.c.})\otimes\sigma_{y}
\end{equation}
for $t\in\left[\ell+\frac{1}{3},\ell+\frac{2}{3}\right)$, and
\begin{alignat}{1}
\hat{{\cal H}}=\hat{{\cal H}}_{3}= & \frac{3J_{3}}{2}\sum_{n_{x},n_{y}}(M|n_{x},n_{y}\rangle\langle n_{x},n_{y}|\\
+ & |n_{x},n_{y}\rangle\langle n_{x}+1,n_{y}|+|n_{x},n_{y}\rangle\langle n_{x},n_{y}+1|+{\rm h.c.})\otimes\sigma_{z}\nonumber 
\end{alignat}
for $t\in\left[\ell+\frac{2}{3},\ell+1\right)$, where $n_{x},n_{y},\ell$
are integers and $\sigma_{x,y,z}$ are Pauli matrices. In
this manuscript we set the Planck constant, driving period,
and lattice constant all equal to $1$.

In the first one third of a driving period, the system is described
by the Hamiltonian $\hat{{\cal H}}_{1}$, where there are only NN
hoppings along $x$-direction of the lattice with a hopping amplitude
$\frac{3J_{1}}{2}$. In the second one third of a driving period,
the system Hamiltonian is switched to $\hat{{\cal H}}_{2}$, where
there are only NN hoppings along $y$-direction of the lattice with
a hopping amplitude $\frac{3J_{2}}{2}$. Finally, in the last one
third of a driving period, the system Hamiltonian is quenched to $\hat{{\cal H}}_{3}$,
where there are NN hoppings along both $x$ and $y$ directions with
equal hopping amplitudes $\frac{3J_{3}}{2}$, together with an onsite
potential $\frac{3J_{3}M}{2}$. Putting together, the Floquet operator
generating the evolution of the system over a complete driving period
is given by $\hat{U}=e^{-i\frac{1}{3}\hat{{\cal H}}_{3}}e^{-i\frac{1}{3}\hat{{\cal H}}_{2}}e^{-i\frac{1}{3}\hat{{\cal H}}_{1}}$.
To simplify the notation, we introduce an ``effective''
Hamiltonian for each step of quenched evolutions as $\hat{H}_{i}=\frac{1}{3}\hat{{\cal H}}_{i}$
($i=1,2,3$). Then the system we are going to study is described by
the Floquet operator
\begin{equation}
\hat{U}=e^{-i\hat{H}_{3}}e^{-i\hat{H}_{2}}e^{-i\hat{H}_{1}}.\label{eq:U}
\end{equation}

Since only NN couplings in the lattice are required in each step of
the quench, our model should be in principle realizable in photonic
setups like those reported in Ref.~\cite{PhotonFTP2}.
In the following section, we will study the bulk Floquet quasienergy
spectrum and Chern numbers of $\hat{U}$ at different hopping parameters
$J_{1}$ or $J_{2}$. We will further demonstrate that by increasing
the value of $J_{1}$ or $J_{2}$, a sequence of topological phase
transitions can be induced, in which the two Floquet bands of $\hat{U}$
exchange their Chern numbers alternatively upon touching at quasienergies $0$ and
$\pi$, realizing the scheme we described in Sec.~\ref{sec:BBC}.

\section{Bulk spectrum and Chern number}\label{sec:SpectrumCN}
In this section, we study the bulk Floquet spectrum and Chern numbers
of our periodically quenched lattice model. For a lattice with $N_{x}\times N_{y}$
unit cells and under periodic boundary conditions (PBC) along both $x$
and $y$ directions, we can perform a Fourier transform $|k_{x},k_{y}\rangle=\frac{1}{\sqrt{N_{x}N_{y}}}\sum_{n_{x}=1}^{N_{x}}\sum_{n_{y}=1}^{N_{y}}e^{i(k_{x}n_{x}+k_{y}n_{y})}|n_{x},n_{y}\rangle$
to find the Floquet operator as $\hat{U}=\sum_{k_{x},k_{y}}|k_{x},k_{y}\rangle U(k_{x},k_{y})\langle k_{x},k_{y}|$,
where $k_{x,y}\in[0,2\pi)$ are two quasimomenta. The Bloch-Floquet
operator $U(k_{x},k_{y})$ has the form
\begin{equation}
U(k_{x},k_{y})=e^{-iH_{3}(k_{x},k_{y})}e^{-iH_{2}(k_{y})}e^{-iH_{1}(k_{x})},\label{eq:Uk}
\end{equation}
with Bloch Hamiltonians
\begin{alignat}{1}
H_{1}(k_{x}) = &\,J_{1}\sin(k_{x})\sigma_{x}\equiv{\cal K}_{1}\sigma_{x},\\
H_{2}(k_{y}) = &\,J_{2}\sin(k_{y})\sigma_{y}\equiv{\cal K}_{2}\sigma_{y},\\
H_{3}(k_{x},k_{y}) = &\,J_{3}[M+\cos(k_{x})+\cos(k_{y})]\sigma_{z}\equiv{\cal K}_{3}\sigma_{z}.
\end{alignat}
Note in passing that in the static limit, $H_{1}(k_{x})+H_{2}(k_{y})+H_{3}(k_{x},k_{y})$
describes the paradigmatic Qi-Wu-Zhang (QWZ) model of Chern insulators~\cite{QWZ}.
The QWZ model possesses two topologically nontrivial phases in the
range of $M\in(-2,2)$, separated by a phase transition
at $M=0$. In each of these topological nontrivial phases, there is
only a single pair of chiral edge states traversing the band gap.
As will be shown, our simple quench protocol endowed the QWZ model
with much richer topological phase structures that are unique to Floquet
systems. 

\begin{figure}
	\centering
	\includegraphics[width=.45\textwidth]{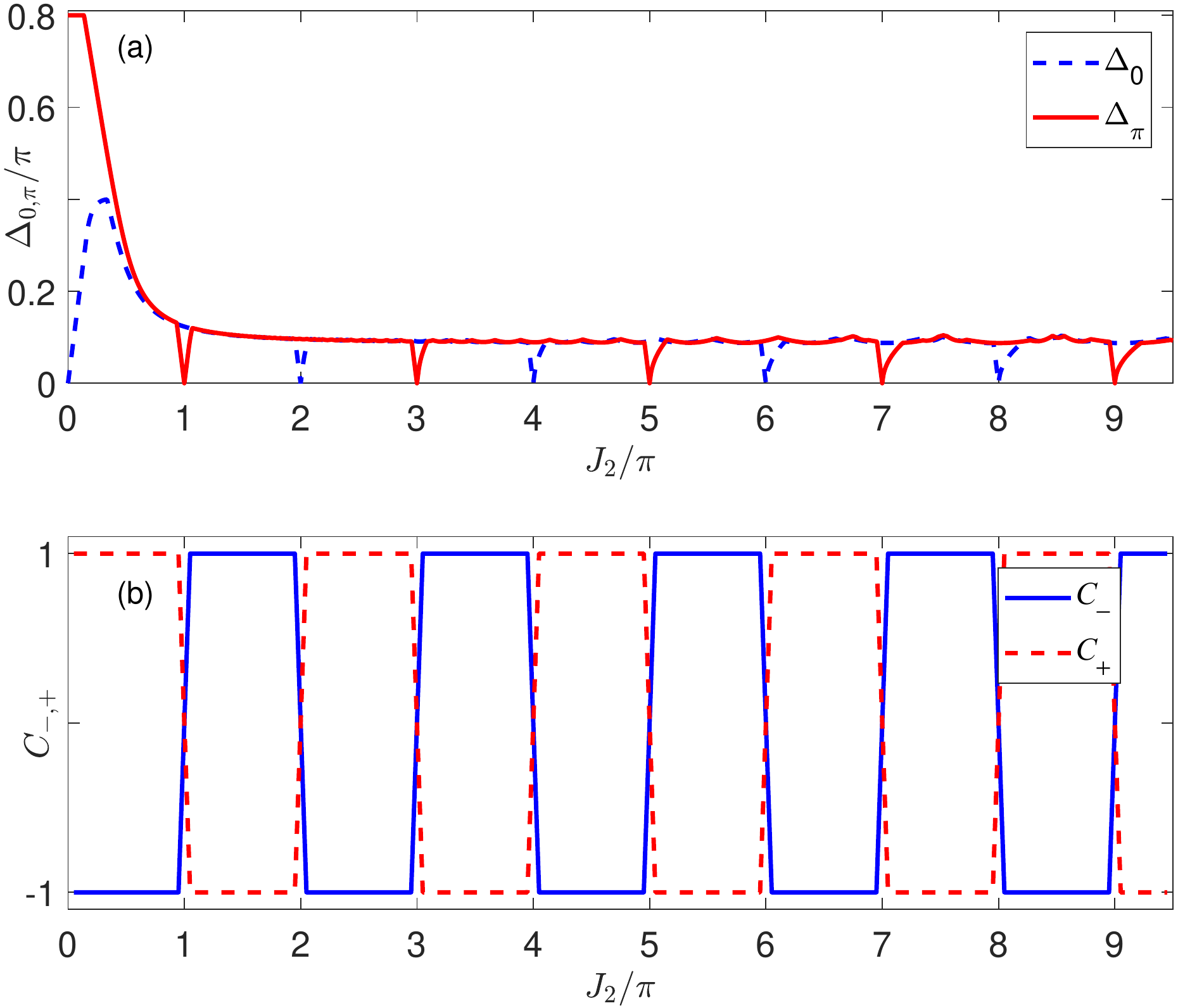}
	\caption{(color online) Floquet spectrum gaps and Chern numbers versus the
		hopping amplitude $J_{2}$ at fixed values of $(J_{1},J_{3},M)=(0.5\pi,0.2\pi,1)$.
		Panel (a): gaps at quasienergies $0$ (blue dashed line) and $\pi$
		(red solid line) as defined in Eqs. (\ref{eq:Gap0}) and (\ref{eq:GapPi}),
		respectively. Panel (b): Chern numbers $C_{-}$ (blue solid line)
		and $C_{+}$ (red dashed line) of the lower and higher Floquet bands $E_-$ and $E_+$,
		respectively.}
	\label{fig:GapCN}
\end{figure}

To see this, let us first check the Floquet spectrum of $U(k_{x},k_{y})$,
which is obtained by solving the eigenvalue equation $U(k_{x},k_{y})|\psi\rangle=e^{-iE(k_{x},k_{y})}|\psi\rangle$.
It is directly seen that the quasienergies (eigenphases) of $U(k_{x},k_{y})$
group into two Floquet bands with dispersions
\begin{alignat}{1}
E_{\pm}(k_{x},k_{y})= & \pm|\arccos[\cos({\cal K}_{3})\cos({\cal K}_{2})\cos({\cal K}_{1})\nonumber\\
+ & \sin({\cal K}_{3})\sin({\cal K}_{2})\sin({\cal K}_{1})]|.\label{eq:Ek}
\end{alignat}
In general, there are two spectrum gaps at quasienergies $0$ and
$\pi$. We characterize them by the gap functions:
\begin{alignat}{1}
\Delta_{0}= & \min_{\{k_{x},k_{y}\}}2|E_{\pm}(k_{x},k_{y})|,\label{eq:Gap0}\\
\Delta_{\pi}= & \min_{\{k_{x},k_{y}\}}2[\pi-|E_{\pm}(k_{x},k_{y})|].\label{eq:GapPi}
\end{alignat}
So the spectrum gap closes at quasienergy $0$ ($\pi$) if $\Delta_{0}=0$
($\Delta_{\pi}=0$). In Fig.~\ref{fig:GapCN}(a), we plot $\Delta_{0}$ (blue dashed
line) and $\Delta_{\pi}$ (red solid line) versus $J_{2}$ at fixed
values of $J_{1}=0.5\pi$, $J_{3}=0.2\pi$ and $M=1$. We observe
that the bulk quasienergy dispersions $E_{\pm}(k_{x},k_{y})$ become gapless
every time when $J_{2}$ hits an integer multiple of $\pi$. Furthermore,
the following pattern of gap closing conditions are identified:
\begin{equation}
J_{2}=\begin{cases}
2n\pi, & \Delta_{0}=0\\
(2n+1)\pi, & \Delta_{\pi}=0
\end{cases}\qquad n=0,1,2,....
\end{equation}
That is, the spectrum gap closes \emph{alternatively} at quasienergies
$0$ and $\pi$ with the increase of $J_{2}$. Similar behaviors of
$\Delta_{0}$ and $\Delta_{\pi}$ versus $J_{1}$ are also found at
fixed values of the other parameters. Also we note that the maximal size
of spectrum gaps at both quasienergies $0$ and $\pi$ is maintained
under the increase of either $J_1$ or $J_2$.

As discussed in Sec.~\ref{sec:BBC}, in order for such a gap evolution process
to generate large winding numbers in both the $0$- and $\pi$-quasienergy
gaps, the two Floquet bands need to exchange their Chern numbers every
time when they touch with each other. To check this, we compute the
Floquet band Chern numbers $C_{\pm}$ of $U(k_{x},k_{y})$ versus
$J_{2}$ at fixed values of $J_{1}=0.5\pi$, $J_{3}=0.2\pi$ and $M=1$.
The Chern number of the Floquet band below (above) quasienergy $0$
is denoted by the blue solid (red dashed) line in Fig.~\ref{fig:GapCN}(b). We see
that the two bands indeed exchange their Chern numbers every time
when $J_{2}$ passes through an integer $(n)$ multiple of $\pi$,
where the gap closes at ($0$-) $\pi$-quasienergy if $n$ is even
(odd). Furthermore, this process happens periodically under the increasing
of $J_{2}$. A similar process is also observed under the increasing
of $J_{1}$ with other system parameters fixed. Putting together,
the quenched lattice model described by Floquet operator (\ref{eq:U})
indeed exemplifies the scheme of generating large gap winding numbers
and chiral edge states as we proposed in Sec.~\ref{sec:BBC}. In the following
section, we will illustrate this point more explicitly by investigating
the spectrum of $\hat{U}$ under open boundary conditions and discussing
its bulk-edge correspondence.

\section{Chiral edge states}\label{sec:EdgeState}
According to the bulk-edge correspondence of $2$d Floquet insulators~\cite{AnomalousESPRX},
the Chern number $C_{\alpha}$ of a bulk Floquet band $\alpha$
can be expressed as 
\begin{equation}
C_{\alpha}=W_{E}[\hat{U}]-W_{E'}[\hat{U}],\label{eq:WCW}
\end{equation}
where $W_{E}$ ($W_{E'}$) is the winding number of the system's Floquet
operator $\hat{U}$ at quasienergy $E$ ($E'$) in the quasienergy
gap above (below) the band $\alpha$. Furthermore, under PBC along
one dimension of the $2$d lattice and OBC along the other, the number
of chiral edge states localized around one edge
of the lattice $n_{{\rm edge}}(E)$ with a quasienergy $E$ in the
gap is related to the winding number as~\cite{AnomalousESPRX}
\begin{equation}
n_{{\rm edge}}(E)=|W_{E}[\hat{U}]|.\label{eq:BER}
\end{equation}

Since a two-band Floquet insulator has two gaps at quasienergies $0$
and $\pi$, the bulk-edge relations~(\ref{eq:WCW}, \ref{eq:BER})
make it possible for the system to have small bulk Chern numbers $C_{\pm}$
but large winding numbers $(W_{0},W_{\pi})$, and therefore many chiral
edge states traversing both of the quasienergy gaps.

The model we introduced in Sec.~\ref{sec:Model} belongs exactly to this situation.
To be explicit, we compute the Floquet spectrum of $\hat{U}$ under
a mixed boundary condition (MBC), for which we denote the case with
OBC/PBC along $x$-direction and PBC/OBC along $y$-direction
of the lattice as MBCX/MBCY. The Floquet operator under MBCX is
denoted by $\hat{U}(k_{y})=e^{-i\hat{H}_{3}(k_{y})}e^{-i\hat{H}_{2}(k_{y})}e^{-i\hat{H}_{1}}$,
where
\begin{alignat}{1}
H_{1}= &\, \frac{J_{1}}{2i}\sum_{n_{x}=1}^{N_{x}-1}(|n_{x}\rangle\langle n_{x}+1|-{\rm h.c.})\otimes\sigma_{x},\\
H_{2}(k_{y})= &\, J_{2}\sin(k_{y})\sum_{n_{x}=1}^{N_{x}}|n_{x}\rangle\langle n_{x}|\otimes\sigma_{y},\\
H_{3}(k_{y})= &\, J_{3}[M+\cos(k_{y})]\sum_{n_{x}=1}^{N_{x}}|n_{x}\rangle\langle n_{x}|\otimes\sigma_{z}\nonumber \\
+ &\, \frac{J_{3}}{2}\sum_{n_{x}=1}^{N_{x}-1}(|n_{x}\rangle\langle n_{x}+1|+{\rm h.c.})\otimes\sigma_{z}.
\end{alignat}
Similarly, the Floquet operator under MBCY is denoted by $\hat{U}(k_{x})=e^{-i\hat{H}_{3}(k_{x})}e^{-i\hat{H}_{2}}e^{-i\hat{H}_{1}(k_{x})}$,
with
\begin{alignat}{1}
H_{1}(k_{x})= &\, J_{1}\sin(k_{x})\sum_{n_{y}=1}^{N_{y}}|n_{y}\rangle\langle n_{y}|\otimes\sigma_{x},\\
H_{2}= &\, \frac{J_{2}}{2i}\sum_{n_{y}=1}^{N_{y}-1}(|n_{y}\rangle\langle n_{y}+1|-{\rm h.c.})\otimes\sigma_{y},\\
H_{3}(k_{x})= &\, J_{3}[M+\cos(k_{x})]\sum_{n_{y}=1}^{N_{y}}|n_{y}\rangle\langle n_{y}|\otimes\sigma_{z}\nonumber \\
+ &\, \frac{J_{3}}{2}\sum_{n_{y}=1}^{N_{y}-1}(|n_{y}\rangle\langle n_{y}+1|+{\rm h.c.})\otimes\sigma_{z}.
\end{alignat}

\begin{figure}
	\centering
	\includegraphics[width=.46\textwidth]{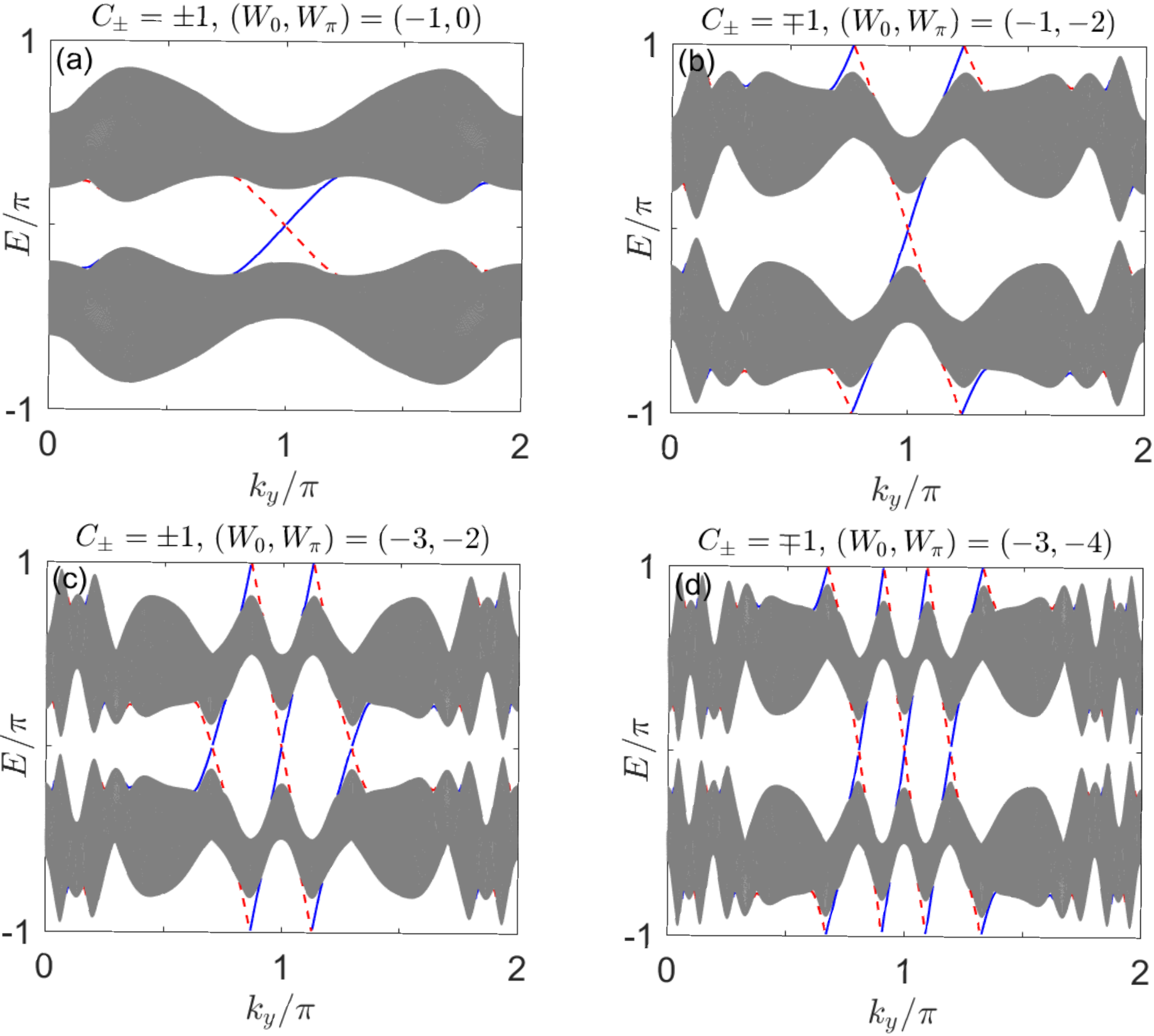}
	\caption{(color online) The Floquet spectrum of $\hat{U}(k_{y})$ under OBC
		and PBC along $x$ and $y$ directions of the lattice, respectively.
		Gray regions represent bulk bands. Blue solid (red dashed) lines
		refer to edge states localized at the left (right) edge of the lattice.
		The lattice has $N_{x}=200$ unit cells along $x$ direction. The
		hopping amplitude $J_{2}$ take values as (a) $J_{2}=0.5\pi$, (b)
		$J_{2}=1.5\pi$, (c) $J_{2}=2.5\pi$ and (d) $J_{2}=3.5\pi$. Other
		system parameters are set at $(J_1,J_3,M)=(0.5\pi,0.2\pi,1)$.
		$C_{-}$ ($C_{+}$) is the Chern numbers of the lower (upper)
		Floquet band and $W_{0}$ ($W_{\pi}$) is the edge state winding number
		at quasienergy $0$ ($\pi$).}
	\label{fig:J2_PBC2OBC1}
\end{figure}

\begin{figure}
	\centering
	\includegraphics[width=.46\textwidth]{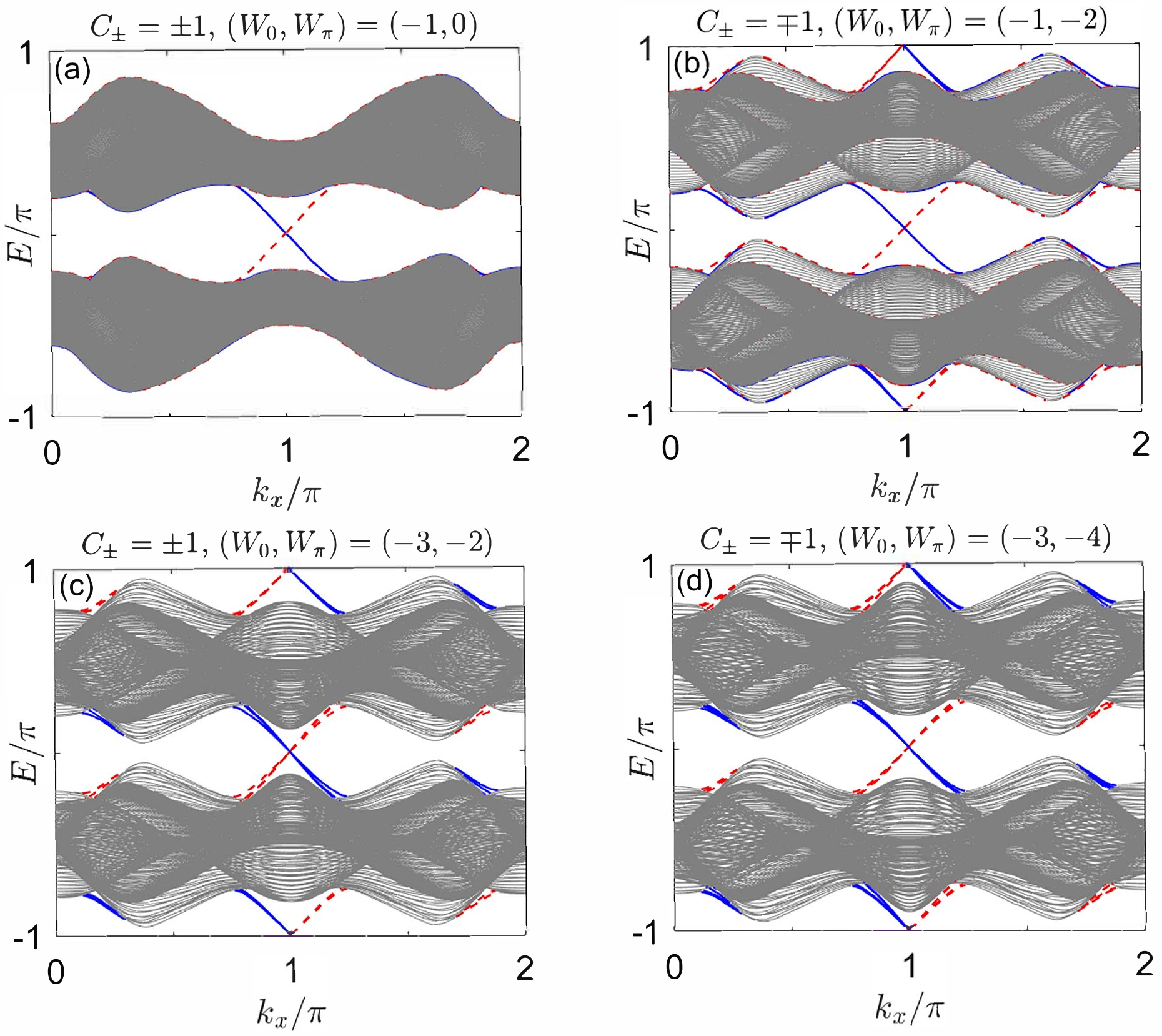}
	\caption{(color online)  The Floquet spectrum of $\hat{U}(k_{x})$ under OBC
		and PBC along $y$ and $x$ directions of the lattice, respectively.
		Gray regions represent bulk bands. Blue solid (red dashed) lines
		refer to edge states localized at the left (right) edge of the lattice.
		The lattice has $N_{y}=200$ unit cells along $y$ direction. The
		hopping amplitude $J_{2}$ take values as (a) $J_{2}=0.5\pi$, (b)
		$J_{2}=1.5\pi$, (c) $J_{2}=2.5\pi$ and (d) $J_{2}=3.5\pi$. Other
		system parameters are set at $(J_1,J_3,M)=(0.5\pi,0.2\pi,1)$.
		$C_{-}$ ($C_{+}$) is the Chern numbers of the lower (upper)
		Floquet band and $W_{0}$ ($W_{\pi}$) is the edge state winding number
		at quasienergy $0$ ($\pi$).}
	\label{fig:J2_PBC1OBC2}
\end{figure}

The quasienergy dispersions of $\hat{U}(k_{y})$ and $\hat{U}(k_{x})$
at several different values of $J_{2}$ are shown in Figs.~\ref{fig:J2_PBC2OBC1} and \ref{fig:J2_PBC1OBC2},
with the number of unit cells $N_{x}=200$ and $N_{y}=200$ for the
two cases, respectively. In all the panels, gray regions represent
bulk Floquet bands and blue solid (red dashed) lines denote chiral
edge states localized around the left (right) boundary of the lattice.
The Chern numbers $C_{\pm}$ of bulk Floquet bands and winding numbers
$W_{0,\pi}$ of chiral edge states at quasienergies $0$ and $\pi$
are also denoted in the figure. The other system parameters are set
at $(J_{1},J_{3},M)=(0.5\pi,0.2\pi,1)$ for all the calculations.

In both Figs.~\ref{fig:J2_PBC2OBC1} and \ref{fig:J2_PBC1OBC2}, we see that two more pairs of chiral edge states
emerge every time when the value of $J_{2}$ increases by $\pi$.
If the gap closes at quasienergy $0$ ($\pi$) during this process,
these new edge states will appear in the gap centered at quasienergy
$0$ ($\pi$) after the transition. Furthermore, for a given $J_{2}$,
the number of chiral edge states in each gap is the same for both
$\hat{U}(k_{y})$ and $\hat{U}(k_{x})$, indicating that these edge
states are insensitive to the configuration of boundary conditions.
More generally, under the chosen set of parameters $(J_{1},J_{3},M)=(0.5\pi,0.2\pi,1)$,
we can infer the following pattern of edge state winding numbers $W_{0,\pi}$
at quasienergies $0$ and $\pi$:
\begin{alignat}{1}
W_{0}= & -2n-1\quad{\rm for}\quad2n\pi<J_{2}<(2n+2)\pi,\\
W_{\pi}= & -2n\qquad\,\,\,{\rm for}\quad(2n-1)\pi<J_{2}<(2n+1)\pi,
\end{alignat}
where $n\in\mathbb{N}$ takes all possible nature numbers. Therefore,
by tuning the value of hopping amplitude $J_{2}$, one can obtain
in principle arbitrarily large winding numbers for both Floquet gaps
centered around quasienergies $0$ and $\pi$. Then according to Eq.
(\ref{eq:BER}), arbitrarily many chiral edge states
could appear in the gaps around quasienergies $0$ and $\pi$. An example
of the spectrum with many chiral edge states is shown in Appendix~\ref{app:EgLarge}.
By varying $J_{1}$ with other system parameters fixed, we observe a similar
pattern for the winding numbers and edge states, with more details presented in
Appendix~\ref{app:UvsJ1}. Two other examples are discussed in Appendix~\ref{app:UvsJ1andJ2}.

Note in passing that in Fig.~\ref{fig:J2_PBC1OBC2}, all the eigenstates at quasienergy
$0$ or $\pi$ in each panel have the same quasimomentum $k_{x}=0$
and also almost the same group velocity $\partial_{k_{x}}E$. Then
for a large enough sample in a large winding number phase, there will
be a significant ``synchronous'' and ``parallel'' topological
current flowing along its edge. Such a current might be more robust
to perturbations and dephasing introduced by the environment, and
therefore has the potential of realizing robust quantum information
transfer.

In both static~\cite{HighCN1,HighCN2,HighCN3,HighCN4,HighCN5,HighCN6,HighCN7} and Floquet~\cite{DerekPRL2012,DerekPRB2014,ZhouEPJB2014,XiongPRB2016,YapPRB2017} $2$d topological insulators, efforts have
been made to engineer bulk bands with large ($>1$) Chern numbers.
An important aim is to find more chiral edge states, and therefore
realizing more quantized and topologically protected transport channels
along the sample edge~\cite{YapPRB2017,YapMajorana2017}.
However, many of the existing approaches require
either more then $2$ bulk bands, or longer range hoppings \emph{plus
}a careful engineering of the local symmetry of the Brillouin zone.
The scheme and quenched lattice model introduced in this manuscript
go around most of these complications, and at the same time allow the
generation of any requested number of chiral edge states in a well
controlled manner. Our results could then serve as a starting point
for both the theoretical exploration of rich Floquet topological phases
in the regime of large winding numbers, and the practical design of
Floquet devices with many quantized edge transport channels.

In the next section, we will demonstrate the transport of chiral edge
states in our system by investigating their two-terminal conductance.
The results show that both the normal and anomalous chiral edge states
give quantized conductance, which are equal to their corresponding
winding numbers.

\section{Two-terminal conductance}\label{sec:Conductance}
In this section, we study the two-terminal transport of chiral edge
states in our quenched lattice model using the approach of Floquet scattering matrix~\cite{Scattering1,Scattering2,Scattering3,Scattering4,Scattering5}.
The $2$d lattice is chosen
to have a patch geometry with OBC along both $x$ and $y$ directions.
The unit cell coordinates $n_{x}$ and $n_{y}$ take values in $1,2,...,N_{x}$
and $1,2,...,N_{y}$, respectively. Two absorbing leads are coupled
to the quenched lattice at its left ($n_{x}=1$) and right ($n_{x}=N_{x}$)
ends. These leads are assumed to act stroboscopically at the start
and end of each Floquet driving period. In the lattice representation,
their effects are described by the following projector onto leads~\cite{Scattering4,Scattering5}:
\begin{equation}
P=\begin{bmatrix}\mathbb{I}_{N_{y}} & \mathbb{O}_{N_{y}\times(N_{x}-1)N_{y}}\\
\mathbb{O}_{N_{y}\times(N_{x}-1)N_{y}} & \mathbb{I}_{N_{y}}
\end{bmatrix}\otimes\sigma_{0},
\end{equation}
where $\mathbb{I}$ ($\mathbb{O}$) represents identity (zero) matrix
and $\sigma_{0}$ is a $2\times2$ identity corresponding to the internal
degrees of freedom. Then for an incoming state with quasienergy $E$
from the left lead to the quenched lattice, we have a fancied scattering
problem described by a quasienergy-dependent scattering matrix~\cite{Scattering4,Scattering5}:
\begin{alignat}{1}
S(E)\equiv & \begin{bmatrix}r(E) & t(E)\\
t'(E) & r'(E)
\end{bmatrix}\\
= & P\left[1-e^{iE}\hat{U}\left(1-P^{T}P\right)\right]^{-1}e^{iE}\hat{U}P^{T},
\end{alignat}
where the Floquet operator $\hat{U}$ is given by Eq. (\ref{eq:U}).
Here the transmission amplitude $t(E)$ is a $2N_{y}$ by $2N_{y}$
matrix, from which the conductance of the quenched lattice (i.e.,
transmission from left to right leads) is obtained as $G(E)={\rm Tr}[t^{\dagger}(E)t(E)]$.
\begin{figure}
	\centering
	\includegraphics[width=.45\textwidth]{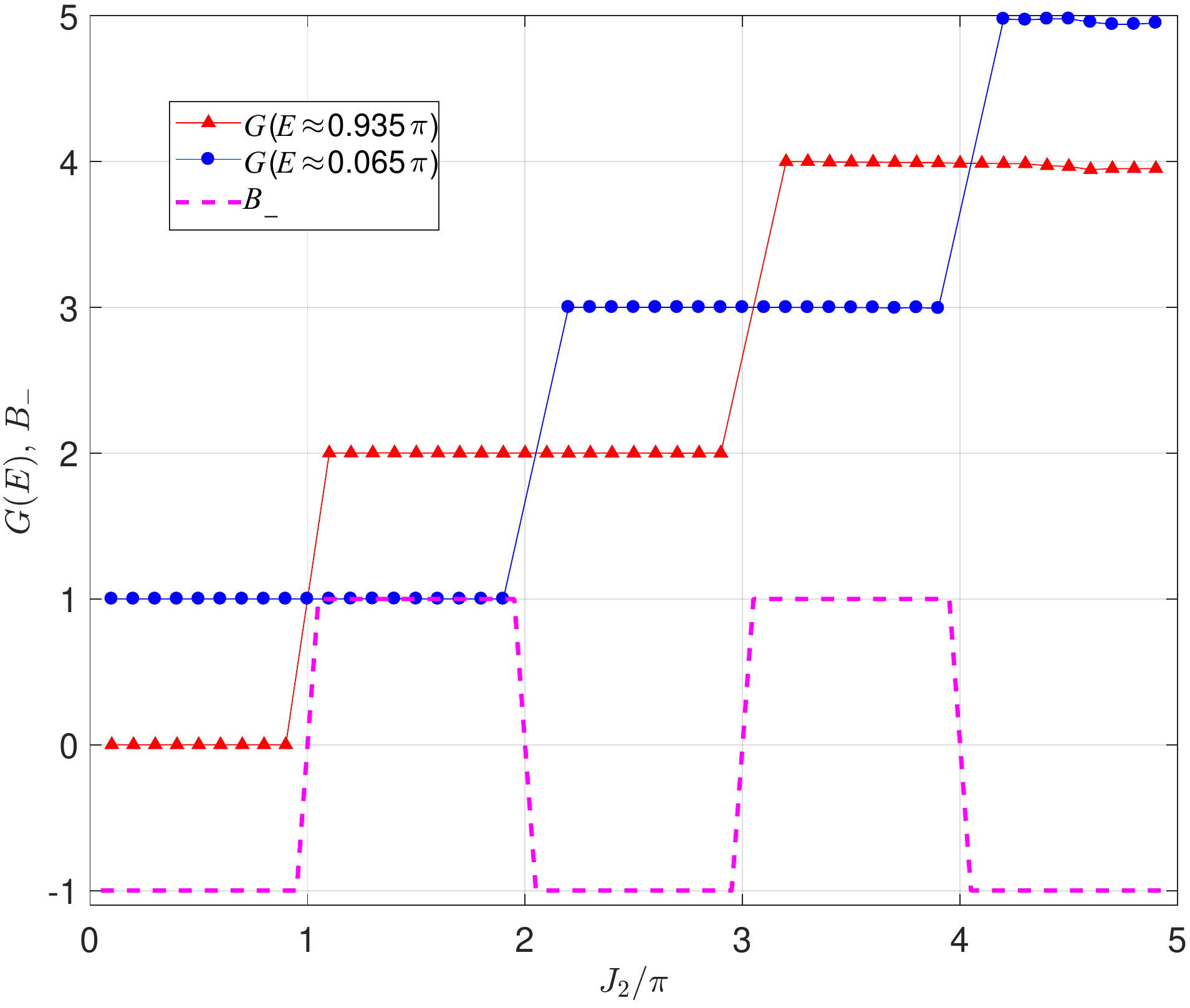}
	\caption{(color online) Two-terminal conductance $G(E)$ and Bott index $B_{-}$
		of $\hat{U}$ versus the hopping amplitude $J_{2}$. Other system
		parameters are fixed at $(J_{1},J_{3},M)=(0.5\pi,0.2\pi,1)$ and the
		lattice size is $N_{x}=N_{y}=70$. Red triangles (blue dots) represent
		the transmission of an incoming state whose quasienergy $E$ is inside
		the Floquet spectrum gap centered around $E=\pi$ ($E=0$). The Bott
		index $B_{-}$ versus $J_{2}$ (magenta dashed line) follows the pattern of the lower band's
		Chern number as shown in Fig.~\ref{fig:Scheme}.}
	\label{fig:J2_Bott}
\end{figure}

In Fig.~\ref{fig:J2_Bott}. we present the calculation of $G(E)$ versus the hopping
amplitude $J_{2}$ at fixed incoming quasienergies $E\approx0$ (blue
dots) and $E\approx\pi$ (red triangles). Referring
to the Chern number pattern and Floquet spectrum presented in Figs.~\ref{fig:GapCN} and \ref{fig:J2_PBC2OBC1}, we clearly see that $G(E)=n_{{\rm edge}}(E)$ for all values
of $J_{2}$ studied here. This further verifies that the
chiral edge states found in our system indeed give quantized conductances
equaling to their winding numbers. For completeness, we also calculated
the Bott index~\cite{TitumPRL2015,Bott1,Bott2,Bott3,Bott4,Bott5,Bott6} of
the lower Floquet band $B_{-}$ in our system (see Appendix~\ref{app:BottCalc} 
for the definition). For a filled Floquet band, the Bott index
is equal to the Chern number, but it is also well-defined in a torus
or patch geometry in position representation. Our results show that
the change of $B_{-}$ versus $J_{2}$ (dashed line in Fig.~\ref{fig:J2_Bott}) follows
exactly the Chern number pattern of the lower band, but is unable to capture
the winding number and the number of chiral edge states traversing
a Floquet gap in our model. This suggests that the winding number
introduced in Ref.~\cite{AnomalousESPRX} might be the most appropriate
invariant to describe topological phases and phase transitions related
to chiral edge states in $2$d Floquet insulators.

Note in passing that in a realistic two-terminal transport setting, an incoming state
is prepared at certain energy instead of quasienergy. In this situation, the quantized edge state conductance is only recovered after applying a ``Floquet sum rule''~\cite{KunduPRL2013}, as also explored in Refs.~\cite{YapPRB2017,YapMajorana2017}.

\section{Summary}\label{sec:summary}
In this manuscript, we proposed a simple Floquet engineering recipe
to generate many topological chiral edge states in a controlled manner.
The essence of our approach is to let the Floquet bands of the system
exchange their Chern numbers periodically and alternatively upon touching at
$0$- and $\pi$-quasienergies. A prototypical quenched lattice
model is introduced to demonstrate our idea. The quantized edge state
conductance of the model in several different topological phases were
obtained from the Floquet scattering matrix of the system. Our results
reveal an intriguing mechanism in the engineering of Floquet transport
devices.

In a realistic system, disorder could have important impacts on its
topology and transport properties~\cite{Disorder1,Disorder2}. The quenched lattice model proposed
in this manuscript could be a promising platform to explore these
effects. On the one hand, the phases with many topological chiral
edge states in our system could be more robust to disorder effects,
and potentially also more efficient in the realization of Floquet
edge state pumps. On the other hand, topological phases with many
chiral edge states are characterized by large winding numbers at both
$0$- and $\pi$-quasienergy gaps. Exploring possible topological
phase transitions induced by disorder in these large winding number
phases is also an interesting topic for future study. 

\section*{Acknowledgement}
J.G. is supported by the Singapore NRF grant No. NRF-NRFI2017-04 (WBS No. R-144-000-378-281) and the Singapore Ministry of Education Academic Research Fund Tier I (WBS No. R-144-000-353-112).

\appendix

\section{Floquet spectrum of $\hat{U}$ with many chiral edge states: an example}\label{app:EgLarge}
\setcounter{equation}{0}
\setcounter{figure}{0}
\numberwithin{equation}{section}
\numberwithin{figure}{section}
In this appendix, we give an example of the Floquet spectrum of $\hat{U}$ defined in Eq.~(\ref{eq:U}) of the main text with many chiral edge states traversing both the gaps around $0$ and $\pi$ quasienergies. The spectrum is shown in Fig.~\ref{fig:J2_PBC2OBC1_Large_WN}, where $9$/$10$ pairs of chiral edge states are found in the spectrum gap centered around quasienergy $0/\pi$. For presentation purpose, only the range of spectrum in which edge states appear is shown.
\begin{figure}
	\centering
	\includegraphics[width=.46\textwidth]{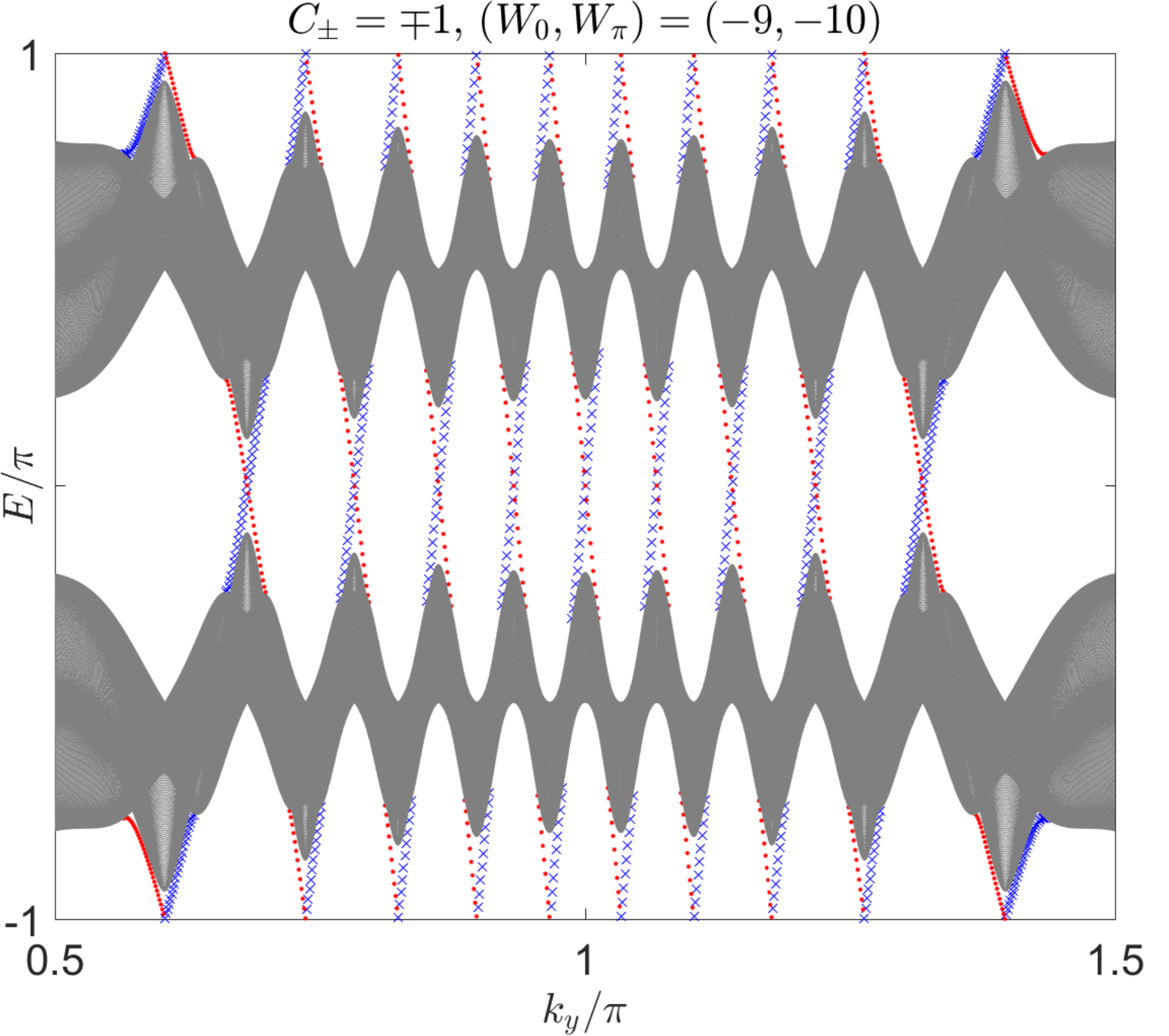}
	\caption{(color online) The Floquet spectrum of $\hat{U}(k_{y})$ under OBC
		and PBC along $x$ and $y$ directions of the lattice, respectively.
		Shaded regions represent bulk bands. Blue crosses (red dots) refer
		to edge states localized at the left (right) edge of the lattice.
		The lattice has $N_{x}=300$ unit cells along $x$ direction. The
		hopping amplitude $J_{2}=9.5\pi$. Other	system parameters are set at $(J_{1},J_{3},M)=(0.5\pi,0.2\pi,1)$. $C_{-}$ ($C_{+}$) is the Chern numbers of the lower (upper)	Floquet band and $W_{0}$ ($W_{\pi}$) is the edge state winding number at quasienergy $0$ ($\pi$).}
	\label{fig:J2_PBC2OBC1_Large_WN}
\end{figure}

\section{Floquet spectrum of $\hat{U}$ at different values of $J_{1}$}\label{app:UvsJ1}
\setcounter{equation}{0}
\setcounter{figure}{0}
\numberwithin{equation}{section}
\numberwithin{figure}{section}

\begin{figure}
	\centering
	\includegraphics[width=.45\textwidth]{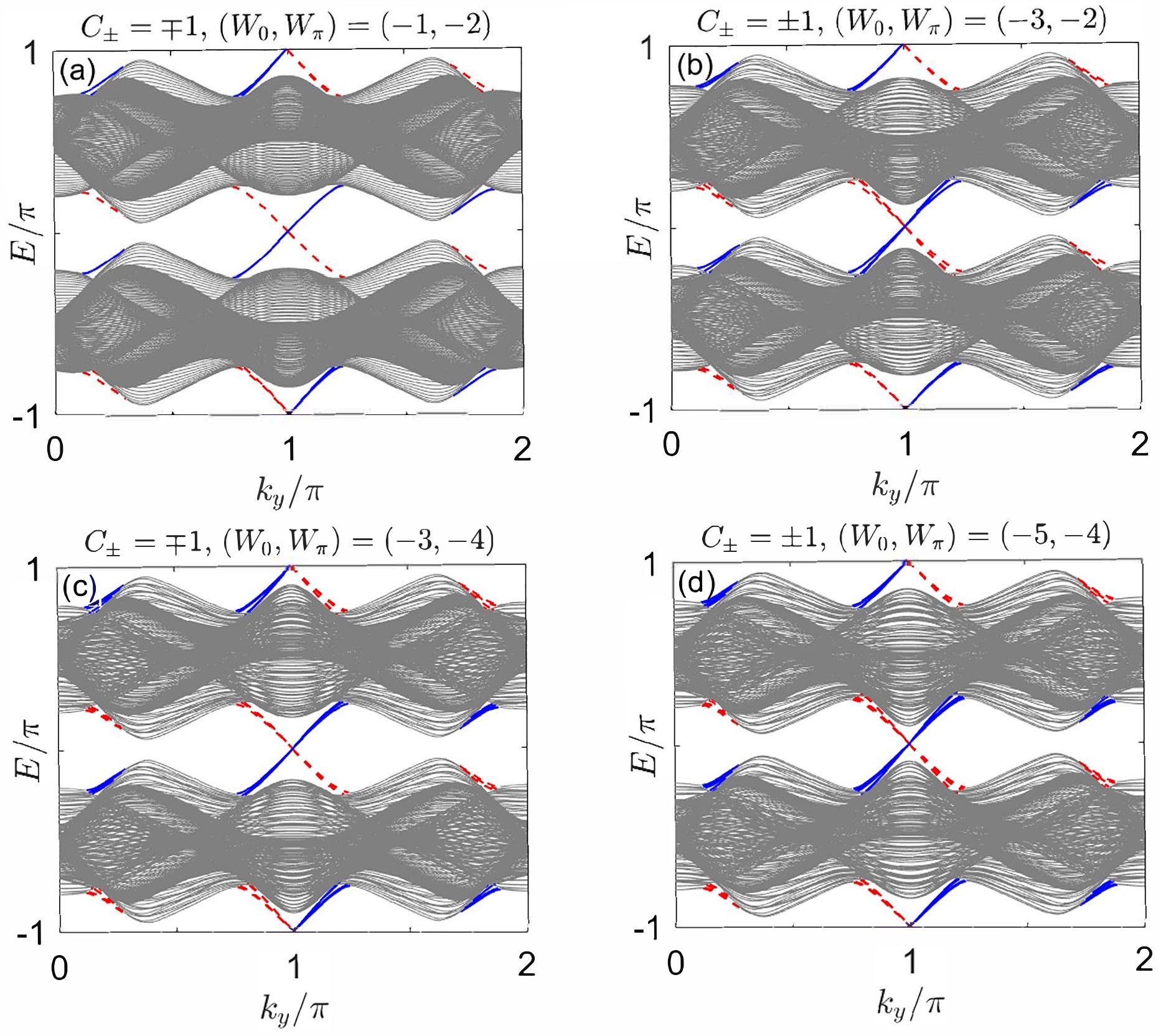}
	\caption{(color online) The Floquet spectrum of $\hat{U}(k_{y})$ under OBC
		and PBC along $x$ and $y$ directions of the lattice, respectively.
		Gray regions represent bulk bands. Blue solid (red dashed) lines
		refer to edge states localized at the left (right) edge of the lattice.
		The lattice has $N_{x}=200$ unit cells along $x$ direction. The
		hopping amplitude $J_{1}$ take values as (a) $J_{1}=1.5\pi$, (b)
		$J_{1}=2.5\pi$, (c) $J_{1}=3.5\pi$ and (d) $J_{1}=4.5\pi$. Other
		system parameters are set at $(J_{2},J_{3},M)=(0.5\pi,0.2\pi,1)$.
		$C_{-}$ ($C_{+}$) is the Chern numbers of the lower (upper)
		Floquet band and $W_{0}$ ($W_{\pi}$) is the edge state winding number
		at quasienergy $0$ ($\pi$).}
	\label{fig:J1_PBC2OBC1}
\end{figure}
\begin{figure}
	\centering
	\includegraphics[width=.45\textwidth]{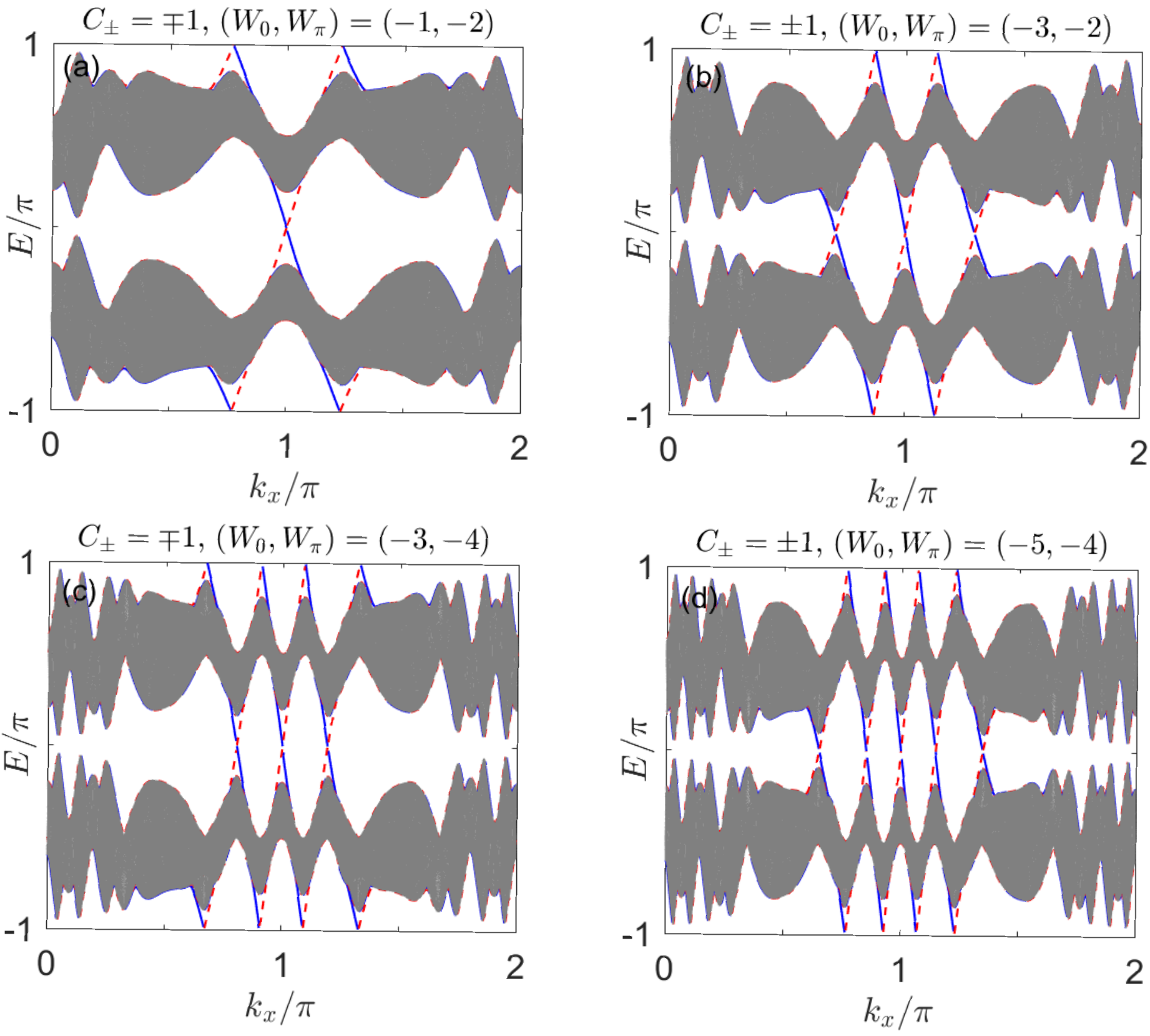}
	\caption{(color online) The Floquet spectrum of $\hat{U}(k_{x})$ under OBC
		and PBC along $y$ and $x$ directions of the lattice, respectively.
		Gray regions represent bulk bands. Blue solid (red dashed) lines
		refer to edge states localized at the left (right) edge of the lattice.
		The lattice has $N_{y}=200$ unit cells along $y$ direction. The
		hopping amplitude $J_{1}$ take values as (a) $J_{1}=1.5\pi$, (b)
		$J_{1}=2.5\pi$, (c) $J_{1}=3.5\pi$ and (d) $J_{1}=4.5\pi$. Other
		system parameters are set at $(J_{2},J_{3},M)=(0.5\pi,0.2\pi,1)$.
		$C_{-}$ ($C_{+}$) is the Chern numbers of the lower (upper)
		Floquet band and $W_{0}$ ($W_{\pi}$) is the edge state winding number
		at quasienergy $0$ ($\pi$).}
	\label{fig:J1_PBC1OBC2}
\end{figure}

In this appendix, we present several more examples of the Floquet spectrum
of $\hat{U}$ versus the hopping amplitude $J_{1}$, with the other
system parameters fixed. Results under MBCX and MBCY are both studied.
We see from Figs.~\ref{fig:J1_PBC2OBC1} and \ref{fig:J1_PBC1OBC2} that the bulk and edge states configurations
are similar to the cases obtained at different values of $J_{2}$
in the main text. This further demonstrate the generality of our Floquet
engineering scheme in the generation of topological phases with large
winding numbers and many chiral edge states.

\section{More examples on the Floquet spectrum}\label{app:UvsJ1andJ2}
\setcounter{equation}{0}
\setcounter{figure}{0}
\numberwithin{equation}{section}
\numberwithin{figure}{section}

\begin{figure}
	\centering
	\includegraphics[width=.45\textwidth]{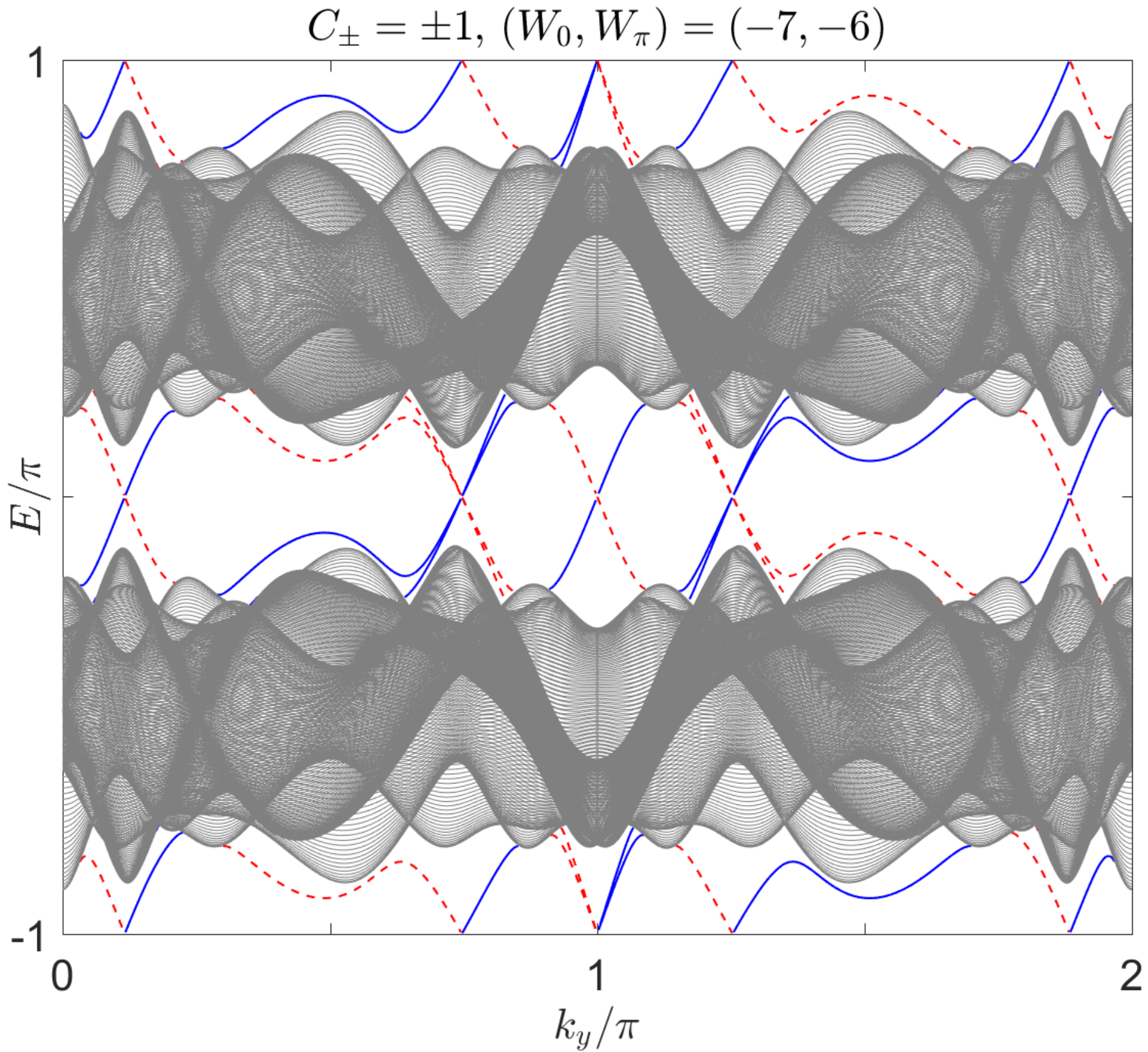}
	\caption{(color online) The Floquet spectrum of $\hat{U}(k_{y})$ under OBC
		and PBC along $x$ and $y$ directions of the lattice, respectively.
		Gray regions represent bulk bands. Blue solid (red dashed) lines
		refer to edge states localized at the left (right) edge of the lattice.
		The lattice has $N_{x}=300$ unit cells along $x$ direction. The
		hopping amplitudes $J_{1}=J_{2}=1.4\pi$. Other
		system parameters are set at $(J_{3},M)=(0.3\pi,1)$.
		$C_{-}$ ($C_{+}$) is the Chern numbers of the lower (upper)
		Floquet band and $W_{0}$ ($W_{\pi}$) is the edge state winding number
		at quasienergy $0$ ($\pi$).}
	\label{fig:J1andJ2_PBC2OBC1_2}
\end{figure}

\begin{figure}
	\centering
	\includegraphics[width=.45\textwidth]{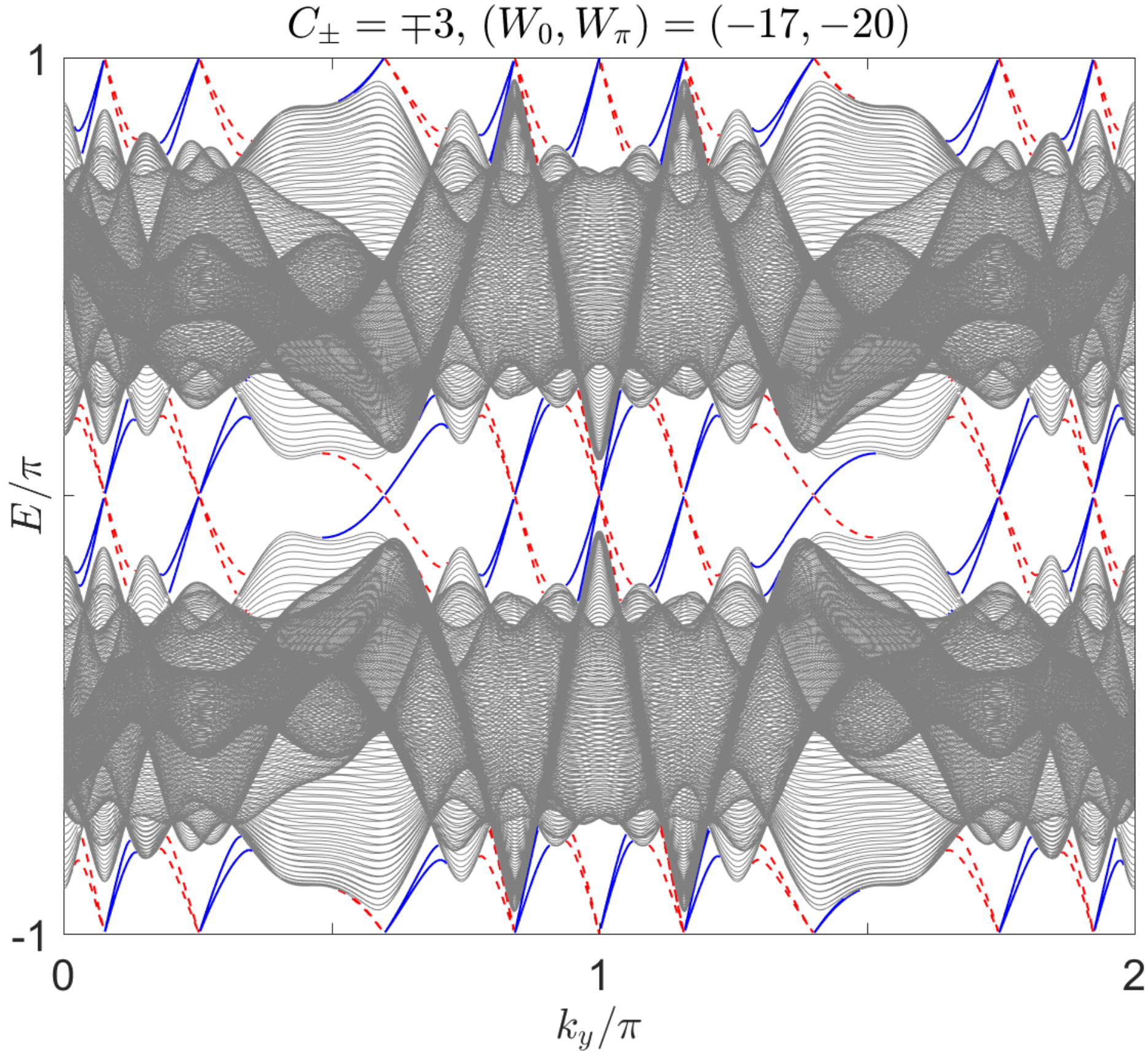}
	\caption{(color online) The Floquet spectrum of $\hat{U}(k_{y})$ under OBC
		and PBC along $x$ and $y$ directions of the lattice, respectively.
		Gray regions represent bulk bands. Blue solid (red dashed) lines
		refer to edge states localized at the left (right) edge of the lattice.
		The lattice has $N_{y}=300$ unit cells along $x$ direction. The
		hopping amplitudes $J_{1}=J_{2}=2.1\pi$. Other
		system parameters are set at $(J_{3},M)=(0.3\pi,1)$.
		$C_{-}$ ($C_{+}$) is the Chern numbers of the lower (upper)
		Floquet band and $W_{0}$ ($W_{\pi}$) is the edge state winding number
		at quasienergy $0$ ($\pi$).}
	\label{fig:J1andJ2_PBC2OBC1_3}
\end{figure}

In this appendix, we present two more examples of the quasienergy spectrum of $\hat{U}(k_{y})$ at different hopping amplitudes $J_1=J_2$, with other system parameters fixed at $(J_3,M)=(0.3\pi,1)$. Numerical results are shown in Figs.~\ref{fig:J1andJ2_PBC2OBC1_2} and \ref{fig:J1andJ2_PBC2OBC1_3}. Similar to the situation in which only one hopping amplitude ($J_1$ or $J_2$) is varied, increasing $J_1$ together with $J_2$ could also induce Chern number exchanges of the two Floquet bands and therefore the growth of the number of chiral edge states in both quasienergy gaps. Furthermore, Floquet bands with Chern numbers larger then $1$ appear in certain parameter windows. However, our numerical calculations suggest that the size of quasienergy gaps will shrink under the joint growth of $J_1$ and $J_2$, accompanied by a more complicated gap closing pattern compared with the one shown in Fig.~\ref{fig:GapCN}(a) of the main text. These make it harder to resolve chiral edge states at larger values of $J_1=J_2$. From another perspective, the complicated topological phase pattern encountered in this situation may call for a statistical analysis of the distribution of winding numbers $(W_0,W_\pi)$ in a wide range of $J_1=J_2$, as considered recently in a one-dimensional system~\cite{Seradjeh2017}.

\section{Calculation of the Bott index}\label{app:BottCalc}
\setcounter{equation}{0}
\setcounter{figure}{0}
\numberwithin{equation}{section}
\numberwithin{figure}{section}

In this appendix, we explain a bit more on the calculation of the Bott
index of our quenched lattice model. Taking a torus geometry of size
$N_{x}\times N_{y}$ for the lattice (i.e., PBC along both $x$ and
$y$ directions), we will have two bulk Floquet bands. We denote $P_{-}$
and $P_{+}$ as projectors to the lower and higher band in the first
quasienergy Brillouin zone, respectively. In the spectrum representation,
these projectors are given by:

\begin{alignat}{1}
P_{-}= & \sum_{E\in(-\pi,0)}E|E\rangle\langle E|=V\begin{bmatrix}\mathbb{I}_{N_{1}N_{2}} & 0\\
0 & 0
\end{bmatrix}V^{\dagger},\\
P_{+}= & \sum_{E\in(0,+\pi)}E|E\rangle\langle E|=V\begin{bmatrix}0 & 0\\
0 & \mathbb{I}_{N_{1}N_{2}}
\end{bmatrix}V^{\dagger},
\end{alignat}
where $V$ is the unitary transformation which diagonalizes the Floquet
operator $\hat{U}$, i.e., $\hat{U}=Ve^{-i\sum_{E}E|E\rangle\langle E|}V^{\dagger}$. 

To evaluate the Bott index, we introduce the ``exponential (unitary)
position operators'' as~\cite{Bott6}:
\begin{alignat}{1}
\hat{U}_{X}= &\, e^{i\frac{2\pi}{N_{x}}\sum_{n_{x}=1}^{N_{x}}n_{x}|n_{x}\rangle\langle n_{x}|\otimes\mathbb{I}_{N_{y}}\otimes\sigma_{0}},\\
\hat{U}_{Y}= &\, e^{i\frac{2\pi}{N_{y}}\mathbb{I}_{N_{x}}\otimes\sum_{n_{y}=1}^{N_{y}}n_{y}|n_{y}\rangle\langle n_{y}|\otimes\sigma_{0}}.
\end{alignat}
Then the projections of these operators to
the lower Floquet band $U_{X-}$ and $U_{Y-}$ are given by
\begin{equation}
P_{-}\hat{U}_{\alpha}P_{-}=V\begin{bmatrix}U_{\alpha-} & 0\\
0 & 0
\end{bmatrix}V^{\dagger},\qquad\alpha=X,Y.
\end{equation}

Finally, the Bott index of the lower Floquet band $B_{-}$ reads
\begin{equation}
B_{-}=\frac{1}{2\pi}{\rm Im}\left\{ {\rm Tr}\left[\ln\left(U_{X-}U_{Y-}U_{X-}^{\dagger}U_{Y-}^{\dagger}\right)\right]\right\} .
\end{equation}
The numerical values of $B_{-}$ for our quenched lattice model are
presented in Fig.~\ref{fig:J2_Bott} of the main text. For a clean sample, it has been
shown that the Bott index of a filled band is equivalent to its Chern
number~\cite{Bott2,Bott5}. However, since the Bott index is defined on a discrete
lattice in position space, it is also well-defined for a disordered
system. Therefore the Bott index could be useful to describe topological
phase transitions induced by disorder in both static and Floquet systems.

\end{document}